\documentclass[conference]{IEEEtran}
\IEEEoverridecommandlockouts
\usepackage{nomencl}
\makenomenclature
\usepackage{amsmath}
\usepackage{etoolbox}
\usepackage{url}

\usepackage{breakurl}
\usepackage[breaklinks]{hyperref}
\usepackage{hyperref}
\usepackage{cite}
\usepackage{float}
\usepackage{graphicx}
\usepackage{tabularx}
\usepackage{fancyhdr}
\allowdisplaybreaks

\makeatletter
\newcommand{\linebreakand}{%
  \end{@IEEEauthorhalign}
  \hfill\mbox{}\par
  \mbox{}\hfill\begin{@IEEEauthorhalign}
}
\makeatother

\author{
  \IEEEauthorblockN{Xinyang Rui}
  \IEEEauthorblockA{\textit{Department of Electrical and Computer Engineering} \\
    \textit{University of Utah}\\
    Salt Lake City, UT, USA \\
    xinyang.rui@utah.edu}
  \and
  \IEEEauthorblockN{Omid Mirzapour}
  \IEEEauthorblockA{\textit{Department of Electrical and Computer Engineering} \\
    \textit{University of Utah}\\
    Salt Lake City, UT, USA \\
    omid.mirzapour@utah.edu}
  \linebreakand % <------------- \and with a line-break
  \IEEEauthorblockN{Brittany Pruneau}
  \IEEEauthorblockA{\textit{Department of Electrical and Computer Engineering} \\
    \textit{University of Utah}\\
    Salt Lake City, UT, USA \\
    u1352355@utah.edu}

    \and
    
  \IEEEauthorblockN{Mostafa Sahraei-Ardakani}
  \IEEEauthorblockA{\textit{Department of Electrical and Computer Engineering} \\
    \textit{University of Utah}\\
    Salt Lake City, UT, USA \\
    mostafa.ardakani@utah.edu}

}
\begin{document}

\title{A Review of Economic Incentives for Efficient Operation of Flexible Transmission\\
\thanks{This research was supported by the National Science Foundation under grant number 2146531.}}

\maketitle
\thispagestyle{plain}
\pagestyle{plain}

\begin{abstract}
The growing penetration of renewable energy requires upgrades to the transmission network to ensure the deliverability of renewable generation. As an efficient alternative to transmission expansion, flexible transmission technologies, whose benefits have been widely studied, can alleviate transmission system congestion and enhance renewable energy integration. However, under the current market structure, investments for these technologies only receive a regulated rate of return, providing little to no incentive for efficient operation. Additionally, a regulated rate of return creates an incentive for building more transmission lines rather than efficient utilization of the existing system. Therefore, investments in flexible transmission technologies remain rather limited. To facilitate the deployment of flexible transmission, improve system efficiency, and accommodate renewable energy integration, a proper incentive structure for flexible transmission technologies, compatible with the current market design, is vital. This paper reviews the current market-based mechanisms for various flexible transmission technologies, including impedance control, dynamic line rating, and transmission switching. This review  pinpoints current challenges of the market-based operation of flexible transmission and provides insights for future endeavors in designing efficient price signals for flexible transmission operation. 
\end{abstract}

%% This code creates the groups
% -----------------------------------------

\renewcommand\nomgroup[1]{%
  \item[\bfseries
  \ifstrequal{#1}{I}{Indices}{%
  \ifstrequal{#1}{S}{Sets}{%
  \ifstrequal{#1}{V}{Variables}{%
  \ifstrequal{#1}{P}{Parameters}{}}}}%
]}
% -----------------------------------------
\mbox{}
\nomenclature[I]{\(n\)}{bus}
\nomenclature[I]{\(k\)}{transmission line}
\nomenclature[I]{\(g\)}{generator}
\nomenclature[I]{\(t\)}{time period}
\nomenclature[S]{\(T\)}{Set of time periods}
\nomenclature[S]{\(l^{+}(n)\)}{Set of lines that are connected "to" bus $n$}
\nomenclature[S]{\(l^{-}(n)\)}{Set of lines that are connected "from" bus $n$}
\nomenclature[S]{\(G\)}{Set of generators}
\nomenclature[S]{\(K\)}{Set of transmission lines}
\nomenclature[S]{\(N\)}{Set of buses}
\nomenclature[S]{\(\mathcal{F}\)}{Set of lines equipped with FACTS, $\mathcal{F} \subset K$}
\nomenclature[V]{\(f_{k}\)}{Active power flow on line $k$}
\nomenclature[S]{\(G(n)\)}{Set of generators connected to bus $n$}
\nomenclature[V]{\(\Delta x_{k}\)}{FACTS reactance adjustment on line $k$}
\nomenclature[V]{\(u_{gt}\)}{Commitment of generator $g$ in period $t$}
\nomenclature[V]{\(\theta_{k, \mathrm{fr}}\)}{Bus voltage angle at the "from" bus of line $k$}
\nomenclature[V]{\(\theta_{k, \mathrm{to}}\)}{Bus voltage angle at the "to" bus of line $k$}
\nomenclature[P]{\(NL_g\)}{No-load cost of generator $g$}
\nomenclature[P]{\(d_{nt}\)}{Demand at bus $n$ in period $t$}
\nomenclature[P]{\(SU_g\)}{Start-up cost of generator $g$}
\nomenclature[P]{\(RU_g\)}{Ramp-up limit of generator $g$}
\nomenclature[P]{\(RD_g\)}{Ramp-down limit of generator $g$}
\nomenclature[P]{\(UT_g\)}{Minimum up time of generator $g$}
\nomenclature[P]{\(DT_g\)}{Minimum down time of generator $g$}
\nomenclature[P]{\(c_g\)}{Linear cost of generator $g$}
\nomenclature[P]{\(f^{\mathrm{max}}_k\)}{Capacity of line $k$}
\nomenclature[P]{\(p^{\mathrm{max}}_g\)}{Maximum generator output of generator $g$}
\nomenclature[P]{\(p^{\mathrm{min}}_g\)}{Minimum generator output of generator $g$}
\nomenclature[P]{\(M\)}{A very large positive number}
\nomenclature[P]{\(\nu_{\mathrm{rated}}\)}{Rated speed of wind turbines}
\nomenclature[P]{\(\nu_{\mathrm{cin}}\)}{Cut-in speed of wind turbines}
\nomenclature[P]{\(\nu_{\mathrm{cout}}\)}{Cut-out speed of wind turbines}
\nomenclature[P]{\(\nu\)}{Wind speed}

\nomenclature[P]{\(\eta_{\mathrm{C}}\)}{Capacitive FACTS compensation level}
\nomenclature[P]{\(\eta_{\mathrm{L}}\)}{Inductive FACTS compensation level}

%\printnomenclature

% \fi
\IEEEpeerreviewmaketitle

\section{Introduction} \label{intro}
\IEEEPARstart{A}{s} the electricity generation and consumption patterns are evolving towards carbon-free generation and electrified consumption worldwide, the transmission system needs upgrades to adapt to the new environment with increased penetration from renewable energy sources (RES). Enhancing renewable integration is essential for decarbonizing the power grid and achieving a carbon-neutral economy, which has been an important objective for countries worldwide. For example, the Biden administration has announced the goal of a net-zero greenhouse gas (GHG) emission economy by 2050~\cite{biden_emission_2050}. Increased levels of renewable energy penetration and growing demand for electrified consumption have led to new congestion patterns in the legacy transmission grid~\cite{navon2020integration}. The geographic locations of renewable resources~\cite{sengupta2018national,panahazari2023hybrid} have added to the congestion in the transmission grid as they can be far away from load centers~\cite{monforti2021impact}.  Therefore, the available transfer capability (ATC) needs enhancement to ensure the deliverability of intermittent and geographically dispersed renewable generation. Transmission expansion is an obvious approach to enhance ATC; however, building new transmission lines faces challenges such as lengthy permitting processes and lumpy investment discouraging investors from investing in this sector~\cite{li2019day,lumbreras2016new}. 

On the other hand, flexible transmission technologies have been viewed by previous literature as an efficient alternative to transmission expansion for ATC procurement and streamlining renewable energ deployment through congestion relief in the transmission system~\cite{gandoman2018review}. Flexible transmission includes a range of technologies and operational methods that allow optimal utilization of current transmission infrastructure instead of considering transmission systems as fixed assets during operation. The adjustments that can provide flexibility in the transmission network include topology changes, reactance compensation, thermal rating adjustment, and nodal phase shift. Prominent flexible transmission technologies include series flexible transmission system (FACTS) devices, transmission switching, dynamic line rating (DLR), high-voltage direct-current (HVDC) lines, and phase-shifting transformers (PST). More detailed descriptions of these technologies are presented in later sections of the paper. An extensive body of literature has shown the potential benefits of implementing flexible transmission to alleviate congestion and facilitate renewable generation integration. 

Despite the widely-studied benefits, flexible transmission deployment in the existing power grid is still limited due to challenges such as the conservative investments in the transmission system, increased computational complexity of operation and planning models with flexible transmission, and lack of economic incentive. An important challenge hindering the implementation of flexible transmission is the lack of a proper market-based incentive structure for transmission assets in current electricity market designs. The most prevalent compensation scheme for transmission investment in several markets does not provide proper incentives for deploying and optimally operating flexible transmission technologies, as the owners receive only a regulated rate of return (RoR) compensation. This is due to the fact that transmission assets were operated as a part of the vertically integrated utilities (VIU) under government regulation. Following the restructuring of power systems, the VIUs disintegrated, and competitive markets were formed for electricity generation and retail. However, the transmission system remained regulated under the umbrella of natural monopoly.  Independent System Operators/Regional Transmission Organizations (ISO/RTOs) were formed to manage wholesale energy and ancillary service markets, operate the transmission network, and plan transmission expansion. Extending the competition to the transmission sector has been a subject of interest since then. The merchant transmission model for compensating transmission investment through Financial Transmission Rights (FTR) has been investigated in several studies~\cite{joskow2005merchant,rubino2015regulatory,staudt2021merchant,biggar2020merchant}. However, under realistic conditions, the benefits of this model are undermined due to stochastic characteristics of the transmission network and market participant behavior~\cite{joskow2005merchant}. With the aforementioned demands for transmission system upgrades due to the steep growth of renewable generation and the need for higher grid resilience, the conventional cost-of-service regulation and monopoly transmission investment projects are insufficient for the changing electricity industry environment. The Federal Energy Regulatory Commission (FERC) Order 1000, issued in 2011, intends to bring competition to US transmission investment by removing barriers, stimulating more participation in transmission investment, and promoting decentralized transmission projects~\cite{fercorder}. Several research endeavors have sought to find optimal investment in flexible transmission technologies in the market environment thereafter~\cite{zhang2018optimal,nourizadeh2022optimal}. A large portion of these efforts have focused on transmission expansion planning with flexible transmission technologies~\cite{tee2016toward, hobbs2016adaptive}

A properly designed market structure facilitates flexible transmission deployment in the deregulated market. Previous literature has proposed different market structures and compensation schemes for flexible transmission. They are based on financial transmission rights or marginal value of flexible transmission operation in day-ahead markets. Regulatory entities and the industry has also pushed for performance-based market structures regarding flexible transmission. Nevertheless, further research is still needed to implement a well-designed market mechanism to harness the benefits of flexible transmission. This paper critically reviews the market structure and incentive proposals for flexible transmission to facilitate further research. 

The rest of this paper is organized as follows: Section~\ref{ft_overview} presents an overview of flexible transmission technologies. Economic valuation and impacts of flexible transmission technologies are presented in Section~\ref{value}, followed by an overview of the proposed market-based incentive structures for flexible transmission operation and concluded by efforts and incentive mechanisms adopted by industry in various ISO/RTOs. The challenges for establishing efficient market mechanisms for flexible transmission are discussed in Section~\ref{challenge}. Finally, conclusions are drawn in Section~\ref{conclusion}, and guidelines for future research are presented. 

\section{Overview Flexible transmission Technologies} \label{ft_overview}
This section presents the functionalities of different flexible transmission technologies in the context of DC power flow. The basic formulation of the single-hour DC optimal power flow (DCOPF) is shown as follows:

\begin{align}
& \min{\sum_{g \in G}c_{g}p_{g}} \label{objopf} \\
& \mathrm{s.t.} \nonumber \\
& p^{\mathrm{min}}_g \leq p_{g} \leq p^{\mathrm{max}}_g, g \in G; \label{genlimopf} \\
& -f^{\mathrm{max}}_k \leq f_{k} \leq f^{\mathrm{max}}_k, k \in K; \label{flowlimopf} \\
& f_{k} = b_k(\theta_{k, \mathrm{to}}-\theta_{k, \mathrm{fr}}), k \in K; \label{floweqnfopf} \\
& \theta_{1} = 0; \label{refbusopf} \\
& \sum_{k \in \delta^{+}(n)}f_{k} - \sum_{k \in \delta^{-}(n)}f_{k} + \sum_{g \in G(n)}p_{g} = d_{n}, n \in N, \label{nbopf}
\end{align}
where $p_g$ is the active power output of generator $g$, $f_k$ is the active power flow through transmission line $k$, and $\theta_{k, \mathrm{to}}$ and $\theta_{k, \mathrm{fr}}$ are the voltage angles at the end buses of line $k$. \eqref{objopf} is the objective function that minimizes the total generation cost, with $c_g$ being the linear marginal cost of generator $g$. Generator capacity limits $p^{\mathrm{max}}_g$ and $p^{\mathrm{min}}_g$ are specified in~\eqref{genlimopf}.~\eqref{flowlimopf} defines the thermal limit constraint of transmission lines and $f^{\mathrm{max}}_k$ is the thermal limit of transmission line $k$. The DC power flow equation, with $b_k$ being the susceptance of transmission line $k$, is presented in~\eqref{floweqnfopf}. \eqref{refbusopf} specifies the voltage angle at the reference bus. Finally,~\eqref{nbopf} is the power balance constraint at each system bus $n$. 

Flexible transmission technologies can enhance operation efficiency and ATC by altering the constraints (\ref{flowlimopf}) and (\ref{floweqnfopf}) in the DCOPF formulation presented above. There are four ways to alter these constraints mathematically: (i) controlling the phase angles in (\ref{floweqnfopf}), (ii) adjusting the susceptance in (\ref{floweqnfopf}), (iii) removing the constraints for a line (switching it out), and (iv) changing the limits in (\ref{flowlimopf}). 

\begin{figure}[t]
\centering
\includegraphics[width=\columnwidth,keepaspectratio]{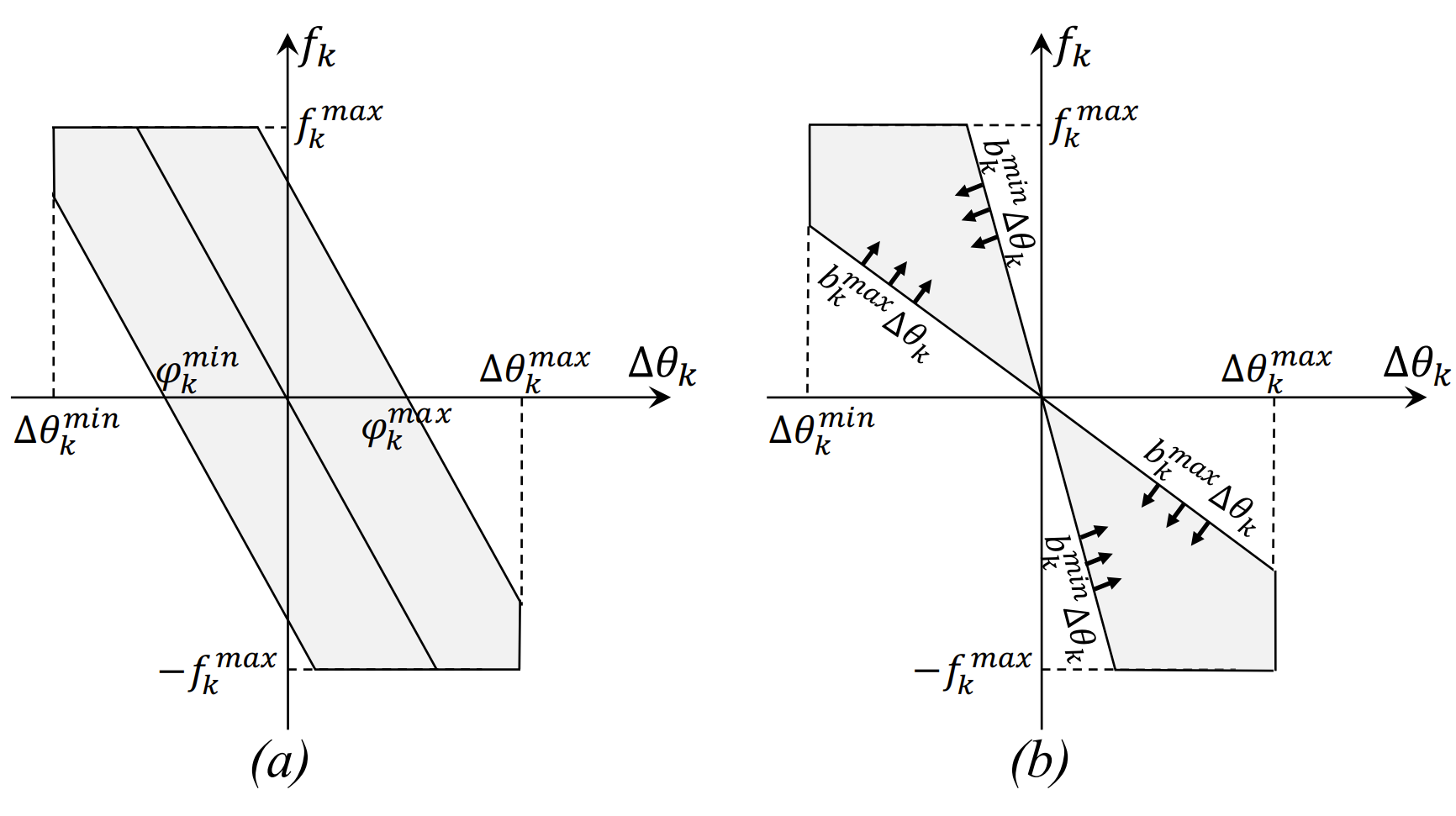}
\caption{Feasible region extension by: (a) phase shift (b) susceptance adjustment.}
\label{feasible}
\end{figure}

PSTs can provide controllability over $\theta_{k, \mathrm{to}}$ and $\theta_{k, \mathrm{to}}$ in~\eqref{floweqnfopf}, effectively enabling the power flow to be controlled \cite{verboomen2005phase}. This controllability can be integrated into the DCOPF formulation by introducing a new variable $\phi_k$ into the line flow constraint~\eqref{floweqnfopf} to extend the feasible region to a wider area. This is shown in Fig.~\ref{feasible}(a).

With the deployment of series FACTS devices, the reactance of transmission lines can be altered so that power flows can be rerouted to avoid transmission bottlenecks. Devices such as the thyristor-controlled series compensator (TCSC), the static synchronous series compensator (SSSC), and the unified power flow controller (UPFC) are widely studied in previous literature and have been deployed in actual industry applications. The TCSC directly adjust the line susceptance, making~\eqref{floweqnfopf} a nonlinear equation. The UPFC and the SSSC use voltage injections to emulate susceptance adjustments. Techniques and modeling to efficiently incorporate series FACTS into power system operation models are presented in~\cite{sahraei2015day, rui2022linear}. The impact of reactance control from TCSC-type devices can be visualized through expanded feasible region as shown in Fig.~\ref{feasible}(b).

With transmission switching, the status of the transmission
elements can be altered so that power flow control functions are provided~\cite{hedman2009optimal}. The formulations of~\eqref{flowlimopf}~and~\eqref{floweqnfopf} are changed, using the big-$M$ method, with the introduction of binary variable $z_k$ representing line switching~\cite{fisher2008optimal}:

\begin{align}
& -f^{\mathrm{max}}_k z_k \leq f_{k} \leq f^{\mathrm{max}}_k z_k, k \in K; \label{flowlimts} \\
& f_{k} - b_k(\theta_{k, \mathrm{to}}-\theta_{k, \mathrm{fr}}) + (z_k - 1)M \leq 0, k \in K; \label{floweqnts1} \\
& f_{k} - b_k(\theta_{k, \mathrm{to}}-\theta_{k, \mathrm{fr}}) - (z_k - 1)M \geq 0, k \in K. \label{floweqnts2}
\end{align}

Transmission switching can also be considered as a discrete susceptance control, where the susceptance is adjusted to zero for a line that is switched out of the system. It is also worth noting that transmission switching can be performed utilizing existing assets~\cite{hedman2011review}, whereas line reactance and phase angle control require the installation of additional devices, which can involve hefty investments.

Under static line rating, the thermal limit $f^{\mathrm{max}}_k$ is a parameter, and traditionally the value is given with a conservative estimate. With DLR, $f^{\mathrm{max}}_k$ is dynamically updated based on monitoring of real-time weather conditions or communication of the actual conductor temperature. Thus, DLR enables the adoption of higher limits that will increase transmission system capacity~\cite{fernandez2016review}.

As an alternative to AC transmission systems, HVDC systems are superior in some applications, including long-distance transmission, offshore renewable integration, and regional electricity market interconnections~\cite{alassi2019hvdc}. The unique controllability features that HVDC systems provide make them suitable for managing congestion and providing flexibility on the grid level~\cite{held2018can}.

\section{Economic Valuation and Market Integration of Flexible Transmission} \label{value}
\subsection{Quantification of Economic Value}
The first step towards designing an efficient market-based scheme for flexible transmission technologies is quantifying the economic benefits of such technologies. This could be evaluated as social welfare enhancement or cost savings in markets with inelastic demand for electricity. Quantifying the benefits is essential for developing market structures to provide the correct incentives for the efficient operation of flexible transmission. A well-designed incentive structure will ensure that the flexible transmission owner benefit is aligned with social welfare improvement. In cases where the optimal direction of adjustment is not aligned with the owner's interest, the market-based scheme should provide compensation schemes for the owner to operate the device in the optimal direction.

The most common benefit of flexible transmission in the existing literature is dispatch cost reduction. Different levels of savings have been reported in various previous studies~\cite{fisher2008optimal, sahraei2015fast, teng2017understanding}. With recent developments in FACTS technology, modular lightweight versions are introduced to the flexible transmission market known as distributed/modular FACTS (D-FACTS or M-FACTS) with enhanced controllability and congestion management capabilities. Ref.~\cite{sang2018economic} evaluates the operation cost savings provided by implementing FACTS and D-FACTS. The evaluation is carried out through a linearized optimal power flow model under different loading scenarios. The results show that the benefits of both FACTS and D-FACTS are higher than break-even costs. D-FACTS offers higher economic value than FACTS incurring cost savings of up to 2.55\%.

\begin{table}[tbh]
    \centering
    \caption{Economic Valuation of Flexible Transmission Technologies}
    \label{tab:flextrans}
    \renewcommand{\arraystretch}{1.2}
    \begin{tabularx}{\columnwidth}{X X X}
        \hline\hline
         Technology & Implementation Cost & Operation Cost Savings\\
         \hline
         FACTS & Medium-High & up to 30\%~\cite{sang2018economic}\\
         TS & Marginal & up to 25\%~\cite{hedman2009optimal,fisher2008optimal,lyon2016harnessing}\\
         HVDC & High & up to 50\%~\cite{alassi2019hvdc,held2018can} \\
         PST & Medium & $\sim$12\%~\cite{konstantelos2014valuation}\\
         DLR & Marginal & $\sim$20\%~\cite{teng2017understanding,uski2015estimation}\\
         \hline\hline
    \end{tabularx}
\end{table}

Other types of benefits are also quantified by previous research. Ref.~\cite{lyon2016harnessing} evaluates the economic benefits of transmission switching providing both congestion alleviation and reliability enhancement in the ISO-NE system. In~\cite{uski2015estimation}, the authors conduct a case study to evaluate the economic benefits of implementing DLR to consumers by studying the electricity prices at both ends of a transmission bottleneck.

Overall, the literature suggests that deploying flexible transmission can provide various benefits, each of which should be quantified separately. These benefits are important in evaluating the performance of flexible transmission technologies and can be the basis for developing economic incentives and compensation mechanisms. The quantification of such benefits is summarized in Table~\ref{tab:flextrans}. This evaluation lays the foundation for efficient incentive design for market operation and further helps the investors with the right choice of technology.

\subsection{Incentive Design and Market Integration}
Although flexible transmission does not possess the same characteristics as bulk transmission expansion projects, they are still regulated as a part of transmission system upgrades and implemented upon ISO/RTO transmission upgrades requirement. This scheme, however, provides no incentive for investing in these technologies, and the deployment has been slow so far. Several compensation mechanisms have been proposed in recent studies to accelerate the proliferation of such technologies through performance-based incentives.
\label{market}

Financial transmission rights (FTRs) are risk-hedging tools designed to minimize the congestion price risk for forward contracts and are successfully implemented in various power markets~\cite{sarkar2008comprehensive}.  In~\cite{sahraei2018merchant}, a market structure, where owners of power flow controllers receive FTR allocations, is proposed to solve the lack of incentive problem under the existing market structure. The authors argued that additional FTRs should be assigned to FACTS device owners. Revenue adequacy and performance of the proposed mechanism are demonstrated on 2-bus and 3-bus systems. However, the difficulty in identifying which particular set of FTRs would correspond to a transmission expansion project, and the order in which projects are built affects the rights awarded can be a drawback for using FTR-style rights to compensate merchant transmission projects~\cite{o2008towards}.

Besides directly using FTR allocation, several previous studies proposed marginal value or other metrics as a compensation mechanism for flexible transmission. It is argued in~\cite{huang2003establishing} that an important issue for using FACTS devices to manage congestion in the deregulated market is the compensation scheme for the utilization of FACTS devices and penalty for users to operate at their limits and addressed both in their proposed price scheme. Under such a scheme, FACTS device owners receive a regulated portion of the total cost savings incurred by their operation. They also receive a penalty from loads when the device is operating at its limits, which is proportional to the value of the Lagrangian dual variable associated with the FACTS operating constraints. However, the proposal is revenue inadequate and incompatible with wholesale energy market structures. Yet, the modeling of FACTS devices presented in this paper is still valuable. Ref.~\cite{sahraei2015transfer} seeks to address the positive externality problem in the transmission payment method proposed in~\cite{o2008towards}.  In \cite{o2008towards}, each transmission element receives a payment equal to the active power flow multiplied by the locational marginal price (LMP) different at the two ends of the element. This creates a positive externality problem, where a flow on a line can increase due to the actions taken by another market player, but the line owner will receive the benefits, due to the increased flow. Ref.~\cite{sahraei2015transfer} alternatively proposes a sensitivity-based calculation of the marginal value of susceptance adjustment by variable-impedance FACTS devices. However, the mathematical proof of revenue adequacy is limited to the case that susceptance adjustment increases the flow on the line that the FACTS device is installed, which is when FACTS increases the absolute value of susceptance. An investment recovery scheme for FACTS devices is proposed in~\cite{mithulananthan2007proposal}, which is based on the load and generator surplus increase due to FACTS deployment. Such a scheme can be utilized as a performance-based incentive for FACTS deployment. In~\cite{fuller2012fast}, similar to the proposal in~\cite{o2008towards}, a metric is introduced to identify the favorable candidate lines for transmission switching. Although this proposal is not intended for the market, it can be used to develop a compensation mechanism. 

Besides the widely studied benefits of reducing the operation cost and facilitating renewable generation integration, the value of flexible transmission in transmission planning as an asset to provide investment flexibility and risk alleviation has also been explored by the existing literature. New transmission projects are capital-intensive and are, in most cases, irreversible. Technologies such as FACTS and PST can provide investment flexibility to avoid unfavorable transmission expansion plans due to uncertainty introduced by future integration of renewable generation~\cite{blanco2011real, konstantelos2014valuation}. Simulation studies in~\cite{blanco2011real} and~\cite{konstantelos2014valuation} show the option value of flexible transmission in long-term transmission expansion projects. The results in~\cite{blanco2011real} show that the option value provided by FACTS devices for deferment of new transmission line investments can be 12\% of the net present value (NPV) of transmission expansion.  It is shown in~\cite{konstantelos2014valuation} that PSTs can bring the total value of \textsterling 13.1 million (reducing investment cost from \textsterling 5609 million to \textsterling 5596 million) in investment cost reduction while providing the transmission expansion planning projects with enough flexibility to reduce uncertainty in investment decisions. However, no existing literature has incorporated such expansion strategies into a merchant transmission scheme. 

\subsection{Industry Practice and ISO/RTO Experience}
\label{industry}
The issues of a regulated RoR and the lack of incentives are known to regulatory entities such as FERC, the ISO/RTOs, and the industry. Over the years, there have been endeavors to make changes and facilitate the deployment of flexible transmission technologies. The U.S. Department of Energy, in a study conducted in the early 2000s, has highlighted the importance of a performance-based regulation (PBR) and revealed that the PBR in the UK led to congestion cost reduction in substantial amounts~\cite{abraham2003national}. Ref.~\cite{abraham2003national} also highlighted that the PBR scheme in the UK showed that incentives for enhanced transmission system operations could have an important role in enhancing transmission operation efficiency, which includes increasing investment in innovative transmission technologies such as flexible transmission. 

Studies regarding the benefits of flexible transmission have been conducted by ISO/RTOs as well. In~\cite{henderson2009planning}, ISO New England (ISO-NE) discussed the value of implementing FACTS and HVDC in their system. Ref.~\cite{henderson2009planning} stated that because of the controllability of HVDC, it is attractive for merchant transmission line applications, and that opportunities for merchant FACTS and HVDC are open in New England. However, no performance-based compensation mechanism regarding merchant transmission projects is mentioned. It is also highlighted by the Pennsylvania-New Jersey-Maryland Interconnection (PJM) that precise control by HVDC makes it ideal for merchant transmission projects~\cite{pjm_hvdc}, with an example being the SOO Green HVDC link~\cite{pjm_SOO}. However, the mechanism is still being developed to incorporate inter-RTO HVDC links into the PJM capacity market to allow customers to benefit from increased competition, greater geographic and technological generation diversity, and the additional instantaneous control offered by dispatchable HVDC facilities~\cite{pjm_SOO}.

In September 2021, FERC held the "Workshop to Discuss Certain Performance-based Ratemaking Approaches"~\cite{fercworkshop}. With a focus on shared savings, this workshop was intended to stimulate the development of transmission technologies. The transmission technologies, or grid-enhancing technologies (GET) discussed at the workshop include flexible transmission technologies such as FACTS devices. The Shared Savings Proposal~\cite{sharedsavings} made by Working for Advanced Transmission Technologies (WATT coalition) and AEE presented a compensation scheme, where 25\% of the savings achieved by implementing transmission technologies are allocated. The proposal also presented a re-evaluation scheme that if the cost-benefit ratio of the project can satisfy the predefined requirement, the incentive will be awarded for the subsequent three years. Despite introducing a performance-based mechanism, this proposal does not have any information regarding making flexible transmission market participants. It also does not address how the compensations should be allocated if multiple projects are planned or carried out in the same time period.  

\section{Challenges and Future Research} \label{challenge}
The existing proposals regarding market structures and incentives for flexible transmission provide important references and guidance for future developments on this topic. The following challenges in this field need to be addressed in future proposals. 
\begin{itemize}
    \item The existing proposals involve a variety of ways of providing incentives/compensations for flexible transmission. They involve shared savings, FTR allocations, as well as generation and load surpluses. Further investigations are needed for each scheme to determine the proper schemes for each technology in different scenarios. A more general compensation scheme for different technologies is desirable. Additionally, Previous studies mainly focused on continuous resources such as FACTS. More compensation mechanism proposals for other technologies are needed~\cite{li2018grid}. Notably, the discrete changes in the network topology have unpredictable impacts on locational marginal prices (LMP) and might create revenue inadequacy in current FTR markets~\cite{hedman2011optimal}. Creating a market-compatible mechanism for accruing the economic benefits of optimal transmission switching remains an interesting research subject.
    \item Several previous studies use small systems which only have two to three buses to demonstrate the effectiveness of the proposed schemes. Numerical studies on larger systems or using real system data are preferable in future research. 
    \item Mathematical proofs, regarding revenue adequacy and alignment of social welfare improvement with flexible transmission incentives, are key for the future development of market structures for flexible transmission. 
\end{itemize}

\section{Conclusions} \label{conclusion}
This paper reviews the existing efforts to establish market structures and economic incentives to stimulate the adoption of flexible transmission technologies. Considering the technological advancements, the lack of performance-based compensation is a key obstacle to increasing the utilization of flexible transmission. Effective solutions to resolve this problem will be vital for enhancing the efficiency of power system operation, improving utilization of the existing transmission system, and ultimately facilitating renewable generation integration to achieve decarbonization targets. While the existing literature provides directions for future research, the problem remains unsolved. We do not yet have appropriate ways to provide incentives for flexible transmission operations.

% or


% \appendices

% \section{}

% \section*{Acknowledgment}


\ifCLASSOPTIONcaptionsoff
  \newpage
\fi



%\bibliographystyle{IEEEtran}

\ifCLASSOPTIONcaptionsoff
  \newpage
\fi

\bibliographystyle{IEEEtran}
\bibliography{IEEEabrv,refs}

\end{document}